\documentclass[
    ,final            
  ]
  {aipproc}

\usepackage{axodraw,amsmath}

\layoutstyle{8x11double}

\graphicspath{{figs/}}

\input{paperdef}

\begin{document}

\onecolumn

\thispagestyle{empty}
\setcounter{page}{0}
\def\thefootnote{\fnsymbol{footnote}}

\begin{flushright}
\mbox{}
arXiv:0809.2396 [hep-ph]
\end{flushright}

\vspace{1cm}

\begin{center}

{\fontsize{15}{1} 
\sc {\bf Discovery Reach of Charged MSSM Higgs Bosons at CMS}}
\footnote{talk given at the {\em SUSY\,08}, 
June 2008, Seoul, Korea}

\vspace{1cm}

{\sc 
S.~Heinemeyer$^1$%
\footnote{
email: Sven.Heinemeyer@cern.ch
}%
, A.~Nikitenko$^2$%
\footnote{
email: Alexandre.Nikitenko@cern.ch
}%
~and G.~Weiglein$^3$%
\footnote{
email: Georg.Weiglein@durham.ac.uk
}
}

\vspace*{1cm}

$^1$ Instituto de Fisica de Cantabria (CSIC-UC), Santander,  Spain

\vspace*{0.1cm}

$^2$ Imperial College, London, UK; on leave from ITEP, Moscow, Russia

\vspace*{0.1cm}

$^3$ IPPP, University of Durham, Durham DH1~3LE, UK

\end{center}

\vspace*{0.2cm}

\BC {\bf Abstract} \EC
We review the $5\,\si$ discovery contours for the charged MSSM Higgs boson at
the CMS experiment with 30~\ifb\ for the two cases $\MHp < \mt$ and 
$\MHp > \mt$. In order to analyze the search reach  
we combine the latest results for the
CMS experimental sensitivities based on full simulation studies with 
state-of-the-art theoretical predictions of MSSM Higgs-boson production
and decay properties.
Special emphasis is put on the SUSY parameter 
dependence of the $5\,\si$ contours.
The variation of $\mu$ can shift the prospective discovery reach 
in $\tb$ by up to $\De\tb~=~40$.

\def\thefootnote{\arabic{footnote}}
\setcounter{footnote}{0}

\newpage


{
\twocolumn

\title{Discovery Reach of Charged MSSM Higgs Bosons at CMS}


\author{S.~Heinemeyer}{
  address={Instituto de Fisica de Cantabria (CSIC-UC), Santander, Spain}
}

\author{A.~Nikitenko}{
  address={Imperial College, London, UK; on leave from ITEP, Moscow, Russia}
}

\author{G.~Weiglein}{
  address={IPPP, University of Durham, Durham DH1~3LE, UK}
}

\begin{abstract}
We review the $5\,\si$ discovery contours for the charged MSSM Higgs boson at
the CMS experiment with 30~\ifb\ for the two cases $\MHp < \mt$ and 
$\MHp > \mt$. In order to analyze the search reach  
we combine the latest results for the
CMS experimental sensitivities based on full simulation studies with 
state-of-the-art theoretical predictions of MSSM Higgs-boson production
and decay properties.
Special emphasis is put on the SUSY parameter 
dependence of the $5\,\si$ contours.
The variation of $\mu$ can shift the prospective discovery reach 
in $\tb$ by up to $\De\tb~=~40$.

\end{abstract}

\maketitle


\section{INTRODUCTION}

One of the main goals of the LHC is the identification of the mechanism
of electroweak symmetry breaking. The most frequently investigated
models are the Higgs mechanism within the Standard 
Model (SM) and within the Minimal Supersymmetric Standard Model
(MSSM). Contrary to the case of the SM, in the MSSM 
two Higgs doublets are required.
This results in five physical Higgs bosons.
These are the light and heavy $\cp$-even Higgs bosons, $h$
and $H$, the $\cp$-odd Higgs boson, $A$, and the charged Higgs bosons,
$H^\pm$.
The Higgs sector of the MSSM can be specified at lowest
order in terms of the gauge couplings, the ratio of the two Higgs vacuum
expectation values, $\tb \equiv v_2/v_1$, and the mass of the $\cp$-odd
Higgs boson, $\MA$. 
Consequently, the masses of the $\cp$-even neutral and the charged Higgs
bosons as well as their production and decay characteristics are dependent
quantities that can be predicted in terms of the Higgs-sector
parameters, e.g.\ 
$\MHp^2 = \MA^2 + \MW^2$, where $\MW$ denotes the mass of the $W$~boson.
Such tree-level results 
in the MSSM are strongly affected by higher-order corrections, in
particular from the sector of the third generation quarks and squarks,
so that the dependencies on various other MSSM parameters can be
important, see e.g.\ \citere{HiggsReviews} for reviews.

Here we review~\cite{cmsHiggs2} the $5\,\si$ charged MSSM Higgs
discovery contours at the LHC for the two cases $\MHp < \mt$ and 
$\MHp > \mt$ within the $\cp$-conserving 
$\mhmax$~scenario~\cite{benchmark2,benchmark3}.
The results are displayed in the $\MHp$--$\tb$ plane.
The respective LHC analyses are given in \citere{HchargedATLAS} for
ATLAS and in \citeres{lightHexp,heavyHexp} for CMS. 
However, within these analyses the variation with relevant SUSY
parameters as well as possibly relevant loop corrections in the Higgs
production and decay~\cite{benchmark3} have been neglected. 
Earlier analyses can be found in \citere{earlier}.


\section{ANALYSIS}

\begin{figure}[htb!]
\includegraphics[width=0.48\textwidth,height=5cm]{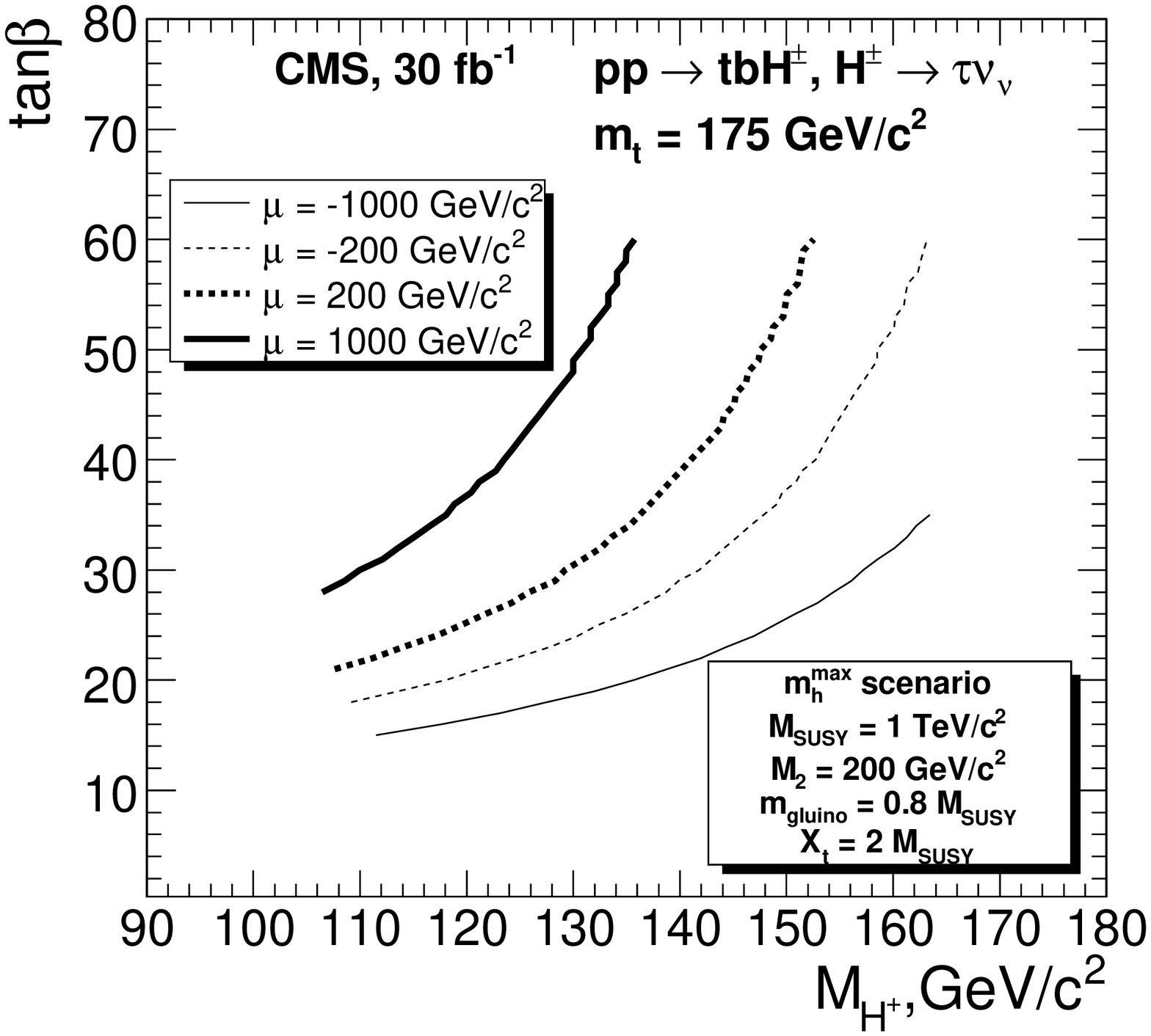}
\includegraphics[width=0.48\textwidth,height=5cm]{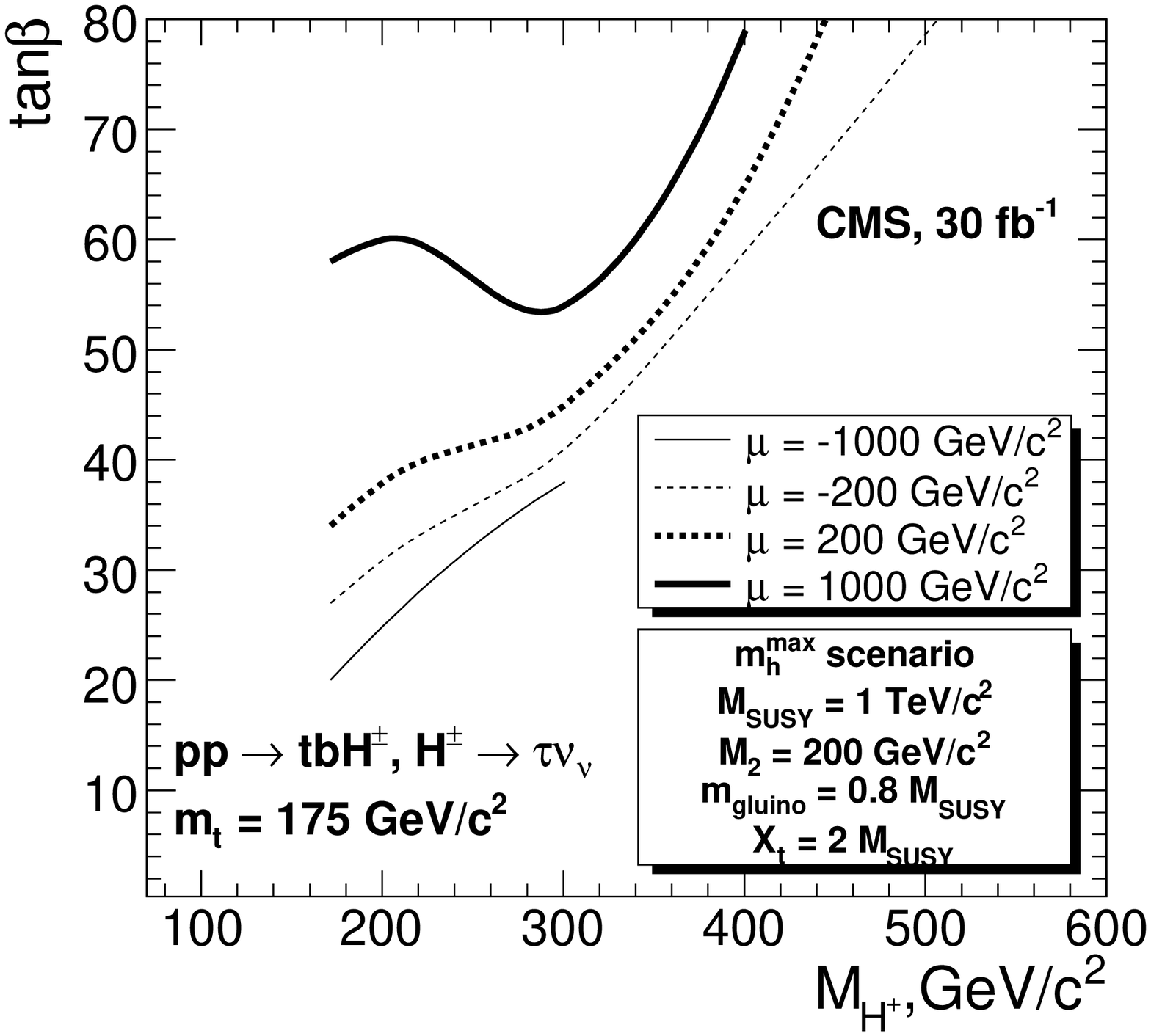}
\caption{Discovery reach for the light (left) and heavy (right) charged
  Higgs boson in the $\MHp$--$\tb$ plane for the $\mhmax$
  scenario~\cite{cmsHiggs2}. 
}
\label{fig:mhmax}
\end{figure}

The analysis of the variation with respect to the relevant SUSY
parameters of the $5\,\si$ discovery contours of the
charged Higgs boson has been performed in \citere{cmsHiggs2}. 
The results have been obtained by using the latest CMS
analyses~\cite{lightHexp,heavyHexp} (based on 30~\ifb) derived in a
model-independent 
approach, i.e.\ making no assumption on the Higgs boson production
mechanism or decays. However, 
only SM backgrounds have been considered.
These experimental results are combined with up-to-date theoretical
predictions for charged Higgs production and decay in the MSSM, taking
into account also the decay to SUSY particles that can in principle
suppress the branching ratio of the charged Higgs boson decay to
$\tau\nu_\tau$.

The main production channels at the LHC are
\begin{align}
\label{pp2Hpm}
pp \to t\bar t \; + \; X, \quad
t \bar t \to t \; H^- \bar b \mbox{~~or~~} H^+ b \; \bar t , \\
gb \to H^- t \mbox{~~or~~} g \bar b \to H^+ \bar t~.
\label{gb2Hpm}
\end{align}
The decay used in the analysis to detect the charged Higgs boson is
\begin{align}
H^\pm \; \to \; \tau \nu_\tau \; \to \; {\rm hadrons~}\nu_\tau. 
\label{Hbug}
\end{align}

The \ul{``light charged Higgs boson''} is characterized by $\MHp < \mt$. 
The main production channel is given in \refeq{pp2Hpm}. Close to
threshold also \refeq{gb2Hpm} contributes. The relevant (i.e.\
detectable) decay channel is given by \refeq{Hbug}.
The experimental analysis is based on 30~\ifb\ collected with CMS.
The events were required to be
selected with the single lepton trigger, thus exploiting the
$W \to \ell \nu$ decay mode of a $W$~boson from the decay of 
one of the top quarks in \refeq{pp2Hpm}.
More details can be found in \citeres{lightHexp,cmsHiggs2}. 

The \ul{``heavy charged Higgs boson''} is characterized by $\MHp \gsim \mt$.
Here \refeq{gb2Hpm} gives the largest contribution to the production cross
section, and very close to 
threshold \refeq{pp2Hpm} can contribute somewhat. The relevant decay
channel is again given in \refeq{Hbug}.
The experimental analysis is based on 30~\ifb\ collected with CMS.
The fully hadronic final state 
topology was considered, thus events were selected with the single
$\tau$ trigger at Level-1 and the combined $\tau$-$E_{\rm T}^{\rm miss}$ High
Level trigger. 
The backgrounds considered were $t \bar t$, $W^\pm t$, 
$W^\pm + 3~{\rm jets}$ as well as  QCD multi-jet
background~\cite{MadGraph,pythia,toprex}. 
The production cross sections for the $t\bar t$~background processes were
normalized to the NLO cross sections~\cite{sigmatt}.
More details can be found in \citeres{heavyHexp,cmsHiggs2}. 

For the calculation of cross sections and branching ratios we use a
combination of up-to-date theory evaluations. The 
interaction of the charged Higgs boson with the $t/b$~doublet can be
expressed in terms of an effective Lagrangian~\cite{deltab2},
\BE
\label{effL}
\cL = \frac{g}{2\MW} \frac{\mbms}{1 + \db} \Bigg[ 
    \wz \, V_{tb} \, \tb \; H^+ \bar{t}_L b_R \Bigg] + {\rm h.c.}
\EE
Here $\mbms$ denotes the running bottom quark mass including SM QCD
corrections. $\db \propto \mu \tb$ depends on the scalar top and bottom
masses, the gluino mass, the Higgs mixing parameter $\mu$ and $\tb$.
The explicit expression can be found in \citeres{deltab1,benchmark3}.

For the production cross section in \refeq{pp2Hpm} we use the SM cross
section $\si(pp \to t \bar t) = 840~\rm{pb}$~\cite{sigmatt}
times the $\br(t \to H^\pm\, b)$ including the $\db$ corrections
described above. 
The production cross section in \refeq{gb2Hpm} is evaluated as given in
\citere{HpmXS}. In addition also the $\db$ corrections of
\refeq{effL} are applied. Finally the $\br(H^\pm \to \tau \nu_\tau)$ is
evaluated taking into account all decay channels, among which the most
relevant are $H^\pm \to tb, cs, W^{(*)}h$. Also possible decays to 
SUSY particles are considered. For the decay to $tb$ again
the $\db$ corrections are included.
All the numerical evaluations are performed with the program 
{\tt FeynHiggs}~\cite{feynhiggs}, see
also \citere{mhcMSSM2L}.


\section{RESULTS}

The numerical analysis has been performed~\cite{cmsHiggs2} in the $\mhmax$
scenario~\cite{benchmark2,benchmark3} for  
$\mu = -1000$, -200, +200, $+1000 \gev$. 
In \reffi{fig:mhmax} we show the results for the variation of the
$5\,\si$ discovery contours for the light (left plot) and the heavy
(right plot) charged Higgs boson, where the charged Higgs boson
discovery will be possible in the areas above the curves shown in the
figure. The top quark mass is set to $\mt = 175 \gev$. 
The thick (thin) lines correspond to positive (negative) $\mu$, and the
solid (dotted) lines have $|\mu| = 1000 (200) \gev$. 

Concerning the light charged Higgs case, 
the curves stop at $\tb = 60$, where we stopped the evaluation of
production cross section and branching ratios. For negative $\mu$ very
large values of $\tb$ result in a strong enhancement of the bottom
Yukawa coupling, and for $\db \to -1$ the MSSM enters a non-perturbative
regime, see \refeq{effL}. 
The search for the light charged Higgs boson covers
the area of large $\tb$ and $\MHp \lsim 130 \ldots 160 \gev$. 
The variation with
$\mu$ induces a strong shift in the $5\,\si$ discovery contours. This
corresponds to a shift in $\tb$ of 
$\De\tb = 15$ for $\MHp \lsim 110 \gev$, rising up to $\De\tb = 40$ for
larger $\MHp$ values. The discovery region is largest (smallest) for 
$\mu = -(+)1000 \gev$, corresponding to the largest (smallest)
production cross section.

We now turn to the heavy charged Higgs case. 
For $\MHp = 170 \gev$, where the experimental
analysis stops, we find a strong variation
in the accessible parameter space for $\mu = -(+)1000 \gev$ of $\De\tb = 40$.
It should be noted in this context that close to threshold, where both
production mechanisms, \refeqs{pp2Hpm} and (\ref{gb2Hpm}), contribute,
the theoretical 
uncertainties are somewhat larger than in the other regions. 
For $\MHp = 300 \gev$ the variation in the $5\,\si$ discovery contours
goes from $\tb = 38$ to $\tb = 54$. For $\mu = -1000 \gev$ and larger
$\tb$ values the bottom Yukawa coupling becomes so large 
that a perturbative treatment would no longer be reliable in this
region, and correspondingly we do not continue the respective curve(s).
Detailed explanations about the shape of the $\mu = +1000 \gev$ curve
for $\MHp \approx 300 \gev$ can be found in \citere{cmsHiggs2}.

\begin{figure}[h!]
\includegraphics[width=0.48\textwidth,height=5cm]{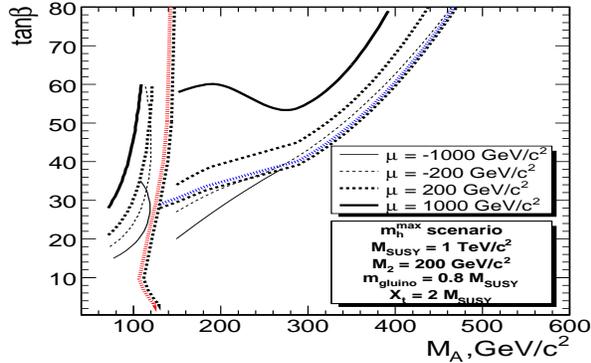}
\caption{
Discovery reach for the charged Higgs boson of CMS with 30~\ifb\ in the 
$\MA$--$\tb$~plane for the $\mhmax$~scenario for 
$\mu = \pm 200, \pm 1000 \gev$ in comparison with the results from the CMS
PTDR~\cite{cmstdr} (see text), obtained for
$\mu = +200 \gev$ and neglecting the $\db$ effects~\cite{cmsHiggs2}. 
}
\label{fig:cmsptdr}
\end{figure}

In \reffi{fig:cmsptdr} we show the 
combined results for the $5\,\si$ discovery contours for the light and
the heavy charged Higgs boson, corresponding to the experimental
analyses in the $\mhmax$ scenario.
They are compared with the results presented in the CMS
PTDR~\cite{cmstdr}, now shown in the $\MA$--$\tb$ plane. 
The thick (thin) lines correspond to positive (negative) $\mu$, and the
solid (dotted) lines have $|\mu| = 1000 (200) \gev$. The thickened
dotted (red/blue) lines represent the CMS PTDR results, obtained for
$\mu = +200 \gev$ and neglecting the $\db$ effects.
Apart from the variation in the $5\,\si$ discovery contours with the
size and the sign of $|\mu|$, two differences can be observed in the
comparison of the PTDR results. 
For the light charged Higgs analysis the discovery contours are now
shifted to smaller $\MA$ values, for negative $\mu$ even ``bending over''
for larger $\tb$ values. The reason is the more complete inclusion of
higher-order corrections to the relation between $\MA$ and $\MHp$ that
is included in {\tt FeynHiggs} as compared to the
calculation used for the CMS PTDR. 
%
The second feature is a small gap between the light and the heavy
charged Higgs analyses, while in the PTDR analysis all charged Higgs
masses could be accessed. 
Possibly the heavy charged Higgs analysis strategy exploiting the fully
hadronic final state can be extended to smaller $\MA$ values to
completely close the gap. 
For the interpretation of \reffi{fig:cmsptdr} it should be kept in mind
that the accessible area in the heavy Higgs analysis also ``bends over''
to smaller $\MA$ values for larger $\tb$, thus decreasing the visible
gap in \reffi{fig:cmsptdr}.
Despite the large LHC reach, the charged Higgs bosons predicted
by recent analyses for simplified SUSY models~\cite{predictions} could
not be covered.


\begin{theacknowledgments}

Work supported in part by the European Community's Marie-Curie Research
Training Network under contract MRTN-CT-2006-035505
`Tools and Precision Calculations for Physics Discoveries at Colliders'.

\end{theacknowledgments}


\bibliographystyle{aipprocl} 

\begin{thebibliography}{99}

\bibitem{HiggsReviews} S.~Heinemeyer, W.~Hollik and G.~Weiglein,
                  {\em Phys.\ Rept.} {\bf 425} (2006) 265;
                  S.~Heinemeyer, 
                  {\em Int. J. Mod. Phys.} {\bf A 21} (2006) 2659;
                  A.~Djouadi,
                  {\em Phys.\ Rept.} {\bf 459} (2008) 1.

\bibitem{cmsHiggs2} M.~Hashemi et al.,
                    arXiv:0804.1228 [hep-ph];

\bibitem{benchmark2} M.~Carena et al.,
                     {\em Eur. Phys. J.} {\bf C 26} (2003) 601.

\bibitem{benchmark3} M.~Carena et al.,
                     {\em Eur.\ Phys.\ J.} {\bf C 45} (2006) 797.

\bibitem{HchargedATLAS} K.~Assamagan, Y.~Coadou and A.~Deandrea,
                        {\em Eur.\ Phys.\ J.\ direct} {\bf C 4} (2002) 9;
                        K.~Assamagan and N.~Gollub,
                        {\em Eur.\ Phys.\ J.} {\bf C 39S2} (2005) 25.

\bibitem{lightHexp} M.~Baarmand, M.~Hashemi and A.~Nikitenko,
                    CMS Note 2006/056.

\bibitem{heavyHexp} R.~Kinnunen,
                    CMS Note 2006/100.

\bibitem{earlier} J.~Coarasa et al.,
                   {\em Eur.\ Phys.\ J.} {\bf C 2} (1998) 373;
                   A.~Belyaev et al.,
                   {\em Phys.\ Rev.} {\bf D 65} (2002) 031701;
                   {\em JHEP} {\bf 0206} (2002) 059;
                   K.~Assamagan et al.,
                   arXiv:hep-ph/0402212;
                   {\em Czech.\ J.\ Phys.} {\bf 55} (2005) B787.

\bibitem{MadGraph}  W.~Long and T.~Stelzer,
                    {\em Comput. Phys. Commun.} {\bf 81} (1994) 357;
                    F.~Maltoni and T.~Stelzer,
                    {\em JHEP} {\bf 0302} (2003) 027.

\bibitem{pythia} T.~Sjostrand et al., 
                 {\em Comput. Phys. Commun.} {\bf 135} (2001) 238.

\bibitem{toprex} S.~Slabospitsky and L.~Sonnenschein,
                 {\em Comput.\ Phys.\ Commun.} {\bf 148} (2002) 87.

\bibitem{sigmatt} P.~Nason, S.~Dawson and R.~K.~Ellis,
                  {\em Nucl.\ Phys.} {\bf B 303} (1988) 607;
                  W.~Beenakker et al.,
                  {\em Phys.\ Rev.} {\bf D 40} (1989) 54;
                  M.~Beneke et al.,
                  arXiv:hep-ph/0003033,
                  and references therein.

\bibitem{deltab2} M.~Carena et al.,
                  {\em Nucl. Phys.} {\bf B 577} (2000) 577.

\bibitem{deltab1} R.~Hempfling,
                  {\em Phys. Rev.} {\bf D 49} (1994) 6168;
                  L.~Hall, R.~Rattazzi and U.~Sarid,
                  {\em Phys. Rev.} {\bf D 50} (1994) 7048;
                  M.~Carena et al.,
                  {\em Nucl. Phys.} {\bf B 426} (1994) 269.

\bibitem{HpmXS} T.~Plehn,
                {\em Phys.\ Rev.} {\bf D 67} (2003) 014018;
                E.~Berger et al.,
                {\em Phys.\ Rev.} {\bf D 71} (2005) 115012.

\bibitem{feynhiggs} S.~Heinemeyer, W.~Hollik and G.~Weiglein,
                    {\em Comput. Phys. Commun.} {\bf 124} (2000) 76;
                     {\em Eur. Phys. J.} {\bf C 9} (1999) 343;
                    G.~Degrassi et al.,
                    {\em Eur. Phys. J.} {\bf C 28} (2003) 133;
                      M.~Frank et al.,
                      {\em JHEP} {\bf 0702} (2007) 047;
                    see: {\tt www.feynhiggs.de} .

\bibitem{mhcMSSM2L} S.~Heinemeyer et al.,
                    {\em Phys.\ Lett.} {\bf B 652} (2007) 300.

\bibitem{cmstdr} CMS Collaboration,
        {\em Physics Technical Design Report, Volume 2. CERN/LHCC
          2006-021}, 
        see: {\tt cmsdoc.cern.ch/cms/cpt/tdr/} .

\bibitem{predictions} O.~Buchmueller et al.,
                      arXiv:0808.4128 [hep-ph];
                      S.~Heinemeyer, M.~Mondrag\'on and G.~Zoupanos,
                      {\em JHEP} {\bf 0807} (2008) 135;
                      arXiv:0809.2397 [hep-ph]];
                      S.~Heinemeyer, 
                      arXiv:0809.2395 [hep-ph]].




\end{thebibliography}


}

\end{document}
